\documentclass[twoside]{article}

\usepackage{graphicx} 
\usepackage{amsfonts} 
\usepackage{amsmath} 

\title{\sc A New Method for tackling Asymmetric Decision Problems}
\author{{\small \bf Peter A. Thwaites}\\
\small School of Mathematics \\ \small University of Leeds\\ \small P.A.Thwaites@leeds.ac.uk
\and
 {\small \bf Jim Q. Smith} \\ \small Department of Statistics\\ \small University of Warwick\\
\small J.Q.Smith@warwick.ac.uk}
\date{}
\begin{document}
\maketitle

\begin{abstract}
Chain Event Graphs are probabilistic graphical models designed especially for the 
analysis of discrete statistical problems which do not admit a natural product space structure. 
We show here how they can be used for decision analysis, and describe an optimal decision strategy based on an efficient local 
computation message passing scheme. We briefly describe a method for producing a parsimonious decision CEG, analogous to
the parsimonious ID, and touch upon the CEG-analogues of Shachter's 
barren node deletion and arc reversal for ID-based solution. 

\end{abstract}

\section{Introduction}

In this paper we demonstrate how the Chain Event Graph (CEG) 
(see for example \cite{PaulandJim,CausalAI,newcap,SilanderCEG,Lorna2}) can be used for 
tackling asymmetric decision problems.

Extensive form (EF) decision trees 
(in which variables appear in the order in which they are observed by a decision maker) are flexible and expressive
enough to represent asymmetries within both the decision and outcome spaces, doing
this through the topological structure of the tree.
They can however become unwieldy, and are not convenient representations from which to read the conditional independence structure of a problem.

Other graphical representations have been developed which to some extent deal with the complexity issue associated with decision
trees, and also allow for local computation. The most commonly used of these is the Influence diagram (ID). 
Because of their popularity, ID solution techniques have developed considerably since their first introduction.
However a major drawback of the ID representation 
is that many decision problems are
asymmetric in that different actions can result in different choices in the future, and IDs are not ideally suited to this sort of problem
\cite{CovandOli}. As decision makers have become more ambitious in the complexity of the problems they wish 
to solve, standard ID and tree-based
methods have proven to be inadequate, and new techniques have become necessary.

There have consequently been many attempts to adapt IDs for use with asymmetric problems (see for example ~\cite{SHM,QZP,JNandShen}), or to 
develop new techniques which use 
both IDs and trees~\cite{CandM}. There have also been several new structures suggested, 
such as Sequential Decision Diagrams (SDDs)~\cite{CovandOli}
and Valuation Networks (VNs)~\cite{Shenoy96}. Asymmetric problems have recently 
also been represented via decision circuits~\cite{BhatandS2}. 
An overview of many of these developments is given by Bielza \& Shenoy in~\cite{BandShen}. They note that none
of the methods available is consistently better than the others. 

CEGs are probabilistic graphical models designed especially for the representation
and analysis of discrete statistical problems which do not admit a natural product space structure. Unlike Bayesian Networks (BNs) 
they are functions of event trees, and
this means that they are able to express the complete sample space structure associated with a problem. They are particularly useful for the analysis
of processes where the future 
development at any specific point depends on the particular history of the problem up to that point. Such dependencies can be thought of as
context-specific conditional independence properties; and the structure implied by these properties
is fully expressed by the topology of the CEG. This is a distinct advantage over context-specific BNs, which
require supplementary information usually in the form of trees or conditional probability tables attached to some of the vertices of the
graph. Like BNs, CEGs provide a suitable framework for efficient local computation algorithms.

Using CEGs for asymmetric decision analysis overcomes drawbacks associated with the current graphs and techniques
used for this purpose. They are an advance on decision trees as they encode the conditional independence structure of problems. They can
represent probability models consistently (which SDDs don't), 
and do not require dummy states or supplementing with extra tables or trees (a drawback of both VNs and Smith et als' adaptations of IDs). 
They can model all asymmetries (which VNs cannot), 
and their semantics are very straightforward, making them an appropriate tool for use by non-experts (both VN \& SDD methodologies are very complicated).

Call \& Miller~\cite{CandM} have drawn attention to the value of coalescence in tree-based approaches to decision problems. They also point out that the difficulties 
in reading conditional independence structure from trees has meant that
analysts using them have not fully taken advantage of the idea of coalescence. They remark that {\it the ability to exploit asymmetry
can be a substantial advantage for trees. If trees could naturally exploit coalescence, the efficiency advantage is even greater}.
SDDs go some way towards exploiting this~\cite{BandShen}, but decision CEGs use coalescence both as a key tool for the expression of conditional 
independence
structure, and to power the analysis.

We show here how CEGs can be used for decision analysis, and describe how to arrive at an optimal decision strategy via an efficient local 
computation message passing scheme. We briefly describe a method for producing a parsimonious decision CEG, analogous to
the parsimonious ID, which contains only those variables and dependencies which the decision maker needs to 
consider when making decisions; and touch upon the CEG-analogues of Shachter's~\cite{Shachter86}
barren node deletion and arc reversal for ID-based solution.

\section{CEGs and decision CEGs}

We start this section with a brief introduction to CEGs -- we direct readers 
to one of~\cite{PaulandJim,newcap} if they would like a more detailed definition. The CEG is a function of a coloured event tree, so 
we begin with a description of these graphs.

\begin{itemize}
\setlength{\itemsep}{-5pt}
\item A coloured event tree $\cal{T}$ is a directed tree with a single root-node. 
\item Each non-leaf-node $v$ has an associated random variable whose state space corresponds to the subset of directed edges of $\cal{T}$ 
	which emanate from~$v$. 
\item Each edge leaving a node $v$ carries a {\it label} which identifies a possible immediate future development given the partial history
	corresponding to the node $v$.
\item The non-leaf-node set of $\cal{T}$ is partitioned into equivalence classes called {\em stages}:
	Nodes in the same {\it stage} have sets of outgoing edges with the same labels, and edges with the same labels have the same probabilities.
\item The edge-set of $\cal{T}$ is partitioned into equivalence classes, whose members have the same {\em colour}:
	Edges have the same {\em colour} when the vertices from which they emanate are in the same stage and the edges have the same label (\& hence
	probability).
\item The non-leaf-node set of $\cal{T}$ is also partitioned into equivalence classes called {\em positions}:
	Nodes are in the same {\em position} if the {\em coloured subtrees} rooted in these nodes are isomorphic both in topology and in colouring
	(so edges in one subtree are coloured (and labelled) identically with their corresponding edges in another). 
\end{itemize}

\noindent{Note that nodes are in the same position 
	when the sets of complete future developments from each node are the same, and have the same probability distribution.}

To produce a CEG $\cal{C}$ from our tree $\cal{T}$, nodes in the same position are combined (as in the coalesced tree),
and all leaf-nodes are combined into a single sink-node. 
We note that for CEGs used for decision problems it is often more convenient to replace the single sink-node by
a set of terminal utility nodes, each of which corresponds to a different utility value. We return to this idea in our example in Section~3.

So the nodes of our CEG $\cal{C}$ are the {\it positions} of the underlying tree $\cal{T}$. We transfer the ideas of {\em stage} and {\em colour} from 
$\cal{T}$ to $\cal{C}$, and
it is this combination of positions and stages that enables the CEG to encode the full conditional independence structure of the problem being
modelled~\cite{PaulandJim}. 

\smallskip
Many discrete statistical processes are asymmetric in that some variables have quite different collections of possible outcomes given different
developments of the process up to that point. It was for these sorts of problem that the CEG was created, and one area where they have proved 
particularly useful is that of causal analysis~\cite{CausalAI,newcap}. 
In much causal analysis the question being asked is {\em If I make this manipulation, what are the effects?}, but graphical models set up to
answer such questions can also be readily used for questions such as {\em If I want to maximise my utility over this process, what are the
manipulations (decisions) I need to make?}

In attempting to answer this second question, we notice that there are only certain nodes or positions in the CEG
which can actually be manipulated. 
We concentrate in this paper on manipulations which impose a probability of one onto one edge emanating from any such node
(equivalent to making a firm decision). Hence the probabilistic nature of these nodes is removed -- they become decision nodes, and we therefore draw 
them as squares.

We draw our CEG in EF order -- as with decision trees this is necessary in order to calculate optimal decision rules.
If two decision nodes in $\cal{T}$ are in the same position, then the optimal strategy is the same for the decision maker (DM) at each of the two
decision nodes: it is conditionally independent of the path taken to
reach the decision node. A similar interpretation can be given to two chance nodes in the same position.

The only other modification that is required to use the CEG for decision analysis is the addition of utilities. This can be done in two ways
(1)~adding utilities to edges, or (2)~expanding the sink-node $w_{\infty}$ into a set of utility nodes, each
corresponding to a distinct utility value (see our example in Section~3). We make our terminal nodes diamond-shaped whether they are leaf nodes
or a single sink-node.

When we manipulate a CEG we prune edges that are given zero probability by the manipulation, and also any edge or position which lies downstream 
of such edges only. No other edges (except those we manipulate to) have probabilities changed by the manipulation \cite{newcap}. This is 
not the case when we simply observe an event, when edge-probabilities upstream of the observation can also change.

In~\cite{DawidinnewCausalbook}, Dawid outlines how a decision-theoretic approach can be taken to causal inference. In this paper we are perhaps
doing the opposite; we show how established causal analysis techniques for CEGs have a natural application in the field of decision analysis.

Our propagation algorithm is illustrated in Table 1 -- at 
the end of the local message passing, the root node will contain the maximum expected utility. 
In the pseudocode we use $C$~\&~$D$ for the sets of chance \& decision nodes,
$p$ represents a probability or weight, and $u$ a utility.  
The utility part of a position~$w$ is denoted by $w[u]$, the probability part of an edge  
by $e(w,w')[p]$ etc. The set of child nodes of a position $w$ is denoted by
$ch(w)$. Note that there may be more than one edge connecting two positions, if say two different decisions have the same consequence. This has
significant ramifications for more complicated problems, as described in our example.

\begin{table}[ht]
\vspace{-4mm}		
\caption{Local propagation algorithm for finding an optimal decision sequence} 
\begin{center}
\begin{flushleft}%
\begin{itemize}
\setlength{\itemsep}{-5pt}
\item Find a topological ordering of the positions. Without loss of generality call this $w_1, w_2,\ldots, w_n$, so that $w_1$ is the
  root-node, and $w_n$ is the sink-node.
\item Initialize the utility value  $w_n[u]$ of the sink node to zero.
\item Iterate: for $i=n-1$ step minus 1 until $i=1$ do:
\begin{itemize}
\item If $w_i \in C$ then $w_i[u] = \sum_{w \in ch(w_i)}\big[\sum_{e(w_i,w)} \big[e(w_i,w)[p]*(w[u]+ e(w_i,w)[u])\big]\big]$
\item If $w_i \in D$ then $w_i[u] = \max_{w \in ch(w_i)}\big[\max_{e(w_i,w)} \big[(w[u]+  e(w_i,w)[u])\big]\big]$ 
\end{itemize}
\item Mark the sub-optimal edges.
\end{itemize}
\end{flushleft}%
\end{center}
\vspace{-4mm}
\end{table}

Note that when we choose to confine utilities to terminal utility nodes, this algorithm is much simplified since both the initializing step
and the $e(w_i,w)[u]$ components are no longer required.

\section{Representing and solving asymmetric decision problems using extensive form CEGs}	

We concentrate here on how the CEG compares with the augmented ID of Smith, Holtzman \& Matheson~\cite{SHM}
for the representation and solution of asymmetric decision problems. We show that the ID-based solution techniques of barren-node 
deletion~\cite{Shachter86} and parsimony 
have direct analogues in the CEG-analysis, and that arc-reversal~\cite{Shachter86} is not 
required for the solution of EF CEGs. The {\it distribution trees}~\cite{SHM} added to the nodes of IDs to describe the 
asymmetry of a problem can simply be thought of as close-ups of interesting parts of the CEG-depiction, where they are an integral part of the 
representation rather than bolt-on as is the case with IDs.
We illustrate this comparison through an example.

We first consider what is meant by conditional independence statements which involve decision variables. 

The statement
$X \amalg Y\ |\ Z$ is true if and only if we can write $P(x\ |\ y,z)$ as $a(x,z)$ for some function $a$ of $x$ and $z$, for all values $x,y,z$ of 
the variables $X,Y,Z$~\cite{Dawid1979}. So clearly, for chance variables $X,Y,Z$ and decision variable $D$, where the value
taken by $X$ is not known to the DM when she makes a decision at $D$,
we can write statements such as $X \amalg D\ |\ Z$ and $X \amalg Y\ |\ D$ since the expressions $P(x\ |\ d,z) = a(x,z)$ and
$P(x\ |\ y,d) = a(x,d)$ are unambiguous in this situation ($d$ representing a value taken by $D$).

Note that $P(d\ |\ y,z)$ is not unambiguously defined, and so conditional independence is no longer a symmetric property when we add decision
variables to the mix.
By a slight abuse of notation we can also write $U \amalg (Y, D_1)\ |\ (Z, D_2)$ if $U(y,z,d_1,d_2) = U(z,d_2)$ for all values $y,z,d_1,d_2$
of the chance variables $Y,Z$ and decision variables $D_1, D_2$.

\smallskip
\noindent{{\bf Example.} {\it Patients suffering from some disease are given one of a set of possible treatments. There is an initial reaction to the treatment in
that the patient's body either accepts the treatment without problems or attempts to reject it. After this initial reaction, the patient responds to
the treatment at some level measurable by their doctor, and this response is independent of the initial reaction conditioned on which treatment
has been given.
The patient's doctor has to make a second decision on how to continue treatment.

There is also the possibility of the patient having some additional condition which affects how they will respond to the treatment. Whether or not they 
have this condition will remain unknown to the doctor, but she can estimate the probability of a patient having it or not (conditioned on their 
response to their particular treatment) from previous studies.

The doctor is concerned with the medium-term health of the patient following her decisions, and knows that this is dependent on whether or not
the patient has the additional condition, how they respond to the first treatment, and the decision made regarding treatment continuation.}

Table 2 summarises this information in the form of a list of variables and relationships.}

\begin{figure}[ht]	
\begin{center}
\includegraphics[height=2.4in]{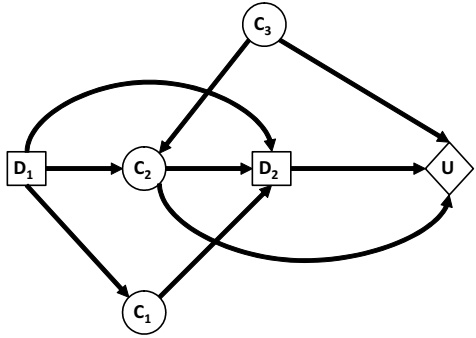}
\caption{EF ID for our example}
\end{center}
\vspace{-8mm}
\end{figure}

\begin{table}[ht]		
\caption{Variables \& relationships; plus $U$ as a function of $C_3, D_2$ and $C_2$} 
       
\begin{center}
\begin{tabular}{lp{20mm}l} 

\begin{tabular}{ll}
$D_1$: &Choice of treatment \\
$C_1$: &Initial reaction \\
$C_2$: &Response to treatment -- $C_2 \amalg C_1\ |\ D_1$ \\
$D_2$: &Decision on how to continue treatment \\
$C_3$: &Condition affecting response to \\
	&treatment and medium-term health \\
     &Can estimate $P(C_3\ |\ D_1,C_2)$ \\
$U$:   &Medium-term health, a function of \\
	&$C_2, D_2$ and $C_3$ \\
\end{tabular}
\hspace{7mm}
\begin{tabular}{ccc|c}
$C_3$ &$C_2$ &$D_2$ &$U$ \\
\hline
1 &1  &1 &$A$ \\
1 &1  &2 &$A$ \\
1 &2  &1 &$B$ \\
1 &2  &2 &$C$ \\
2 &1  &1 &$A$ \\
2 &1  &2 &$A$ \\
2 &2  &1 &$D$ \\
2 &2  &2 &$E$ \\
\end{tabular}
\end{tabular}
\end{center}
\vspace{-5mm}
\end{table}

To avoid making the problem too complex for easy understanding we let all variables be binary except $U$, and introduce only two asymmetric 
features:
So suppose that if a patient fails to respond to the first treatment ($C_2 = 1$), then the patient will inevitably have the lowest medium-term
health rating ($U = A$). We can express this as $U \amalg (C_3, D_2)\ |\ (C_2 = 1)$ (see Table 2).
Suppose also that if $D_1 = 2$ (Treatment 2 is given) then $C_1$ takes the value~1 (the patient's body always accepts the treatment).
The problem can be represented by the EF ID in Figure 1.

To express the asymmetry of the problem we can add {\it distribution trees} to the nodes $C_1$ and $U$ as in Figure~2. These have been drawn in
a manner consistent with the other diagrams in this paper, rather than with those in~\cite{SHM}.

\begin{figure}[ht]	
\includegraphics[height=2.1in]{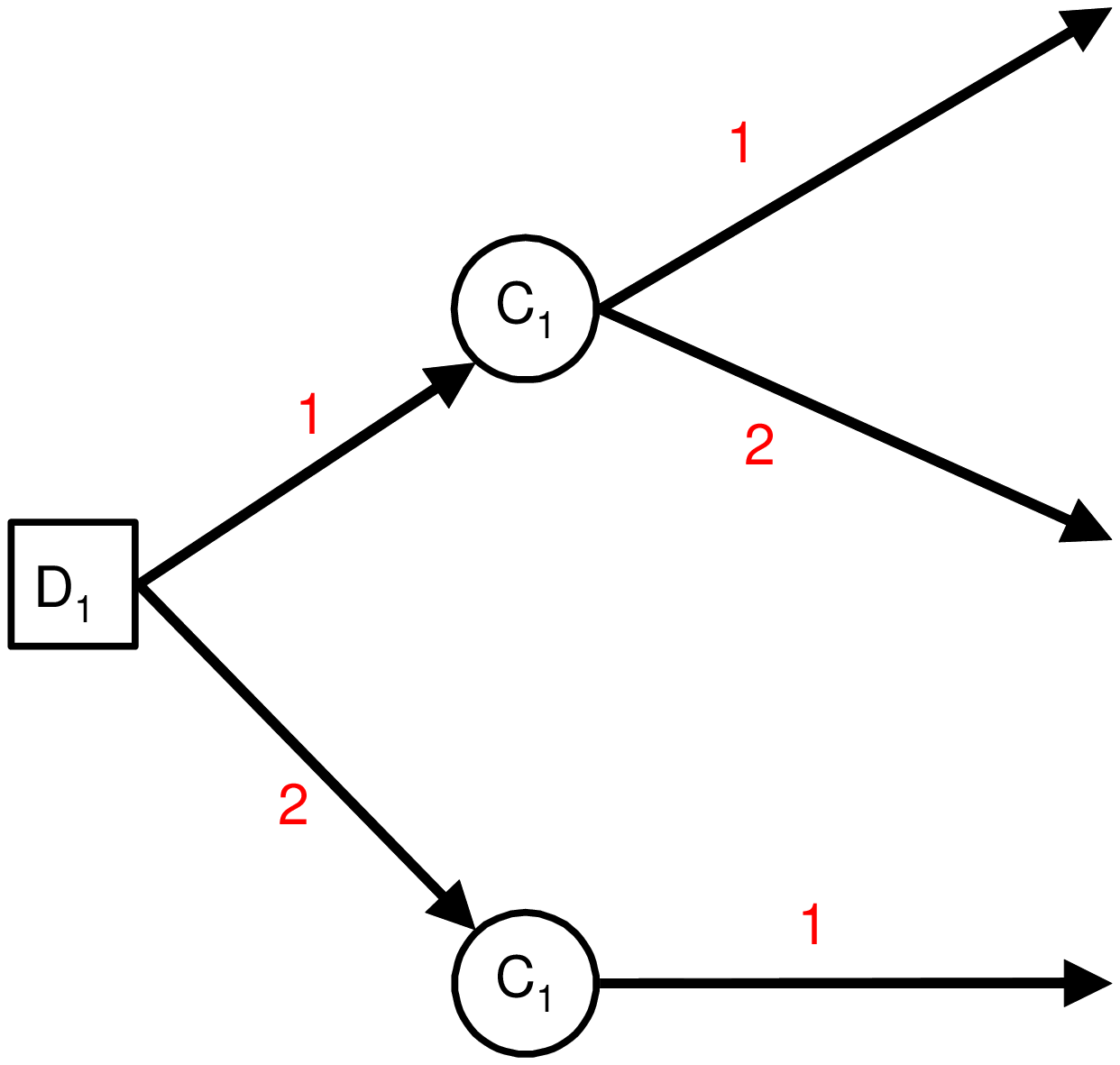}\kern-0.8in\includegraphics[height=2.1in]{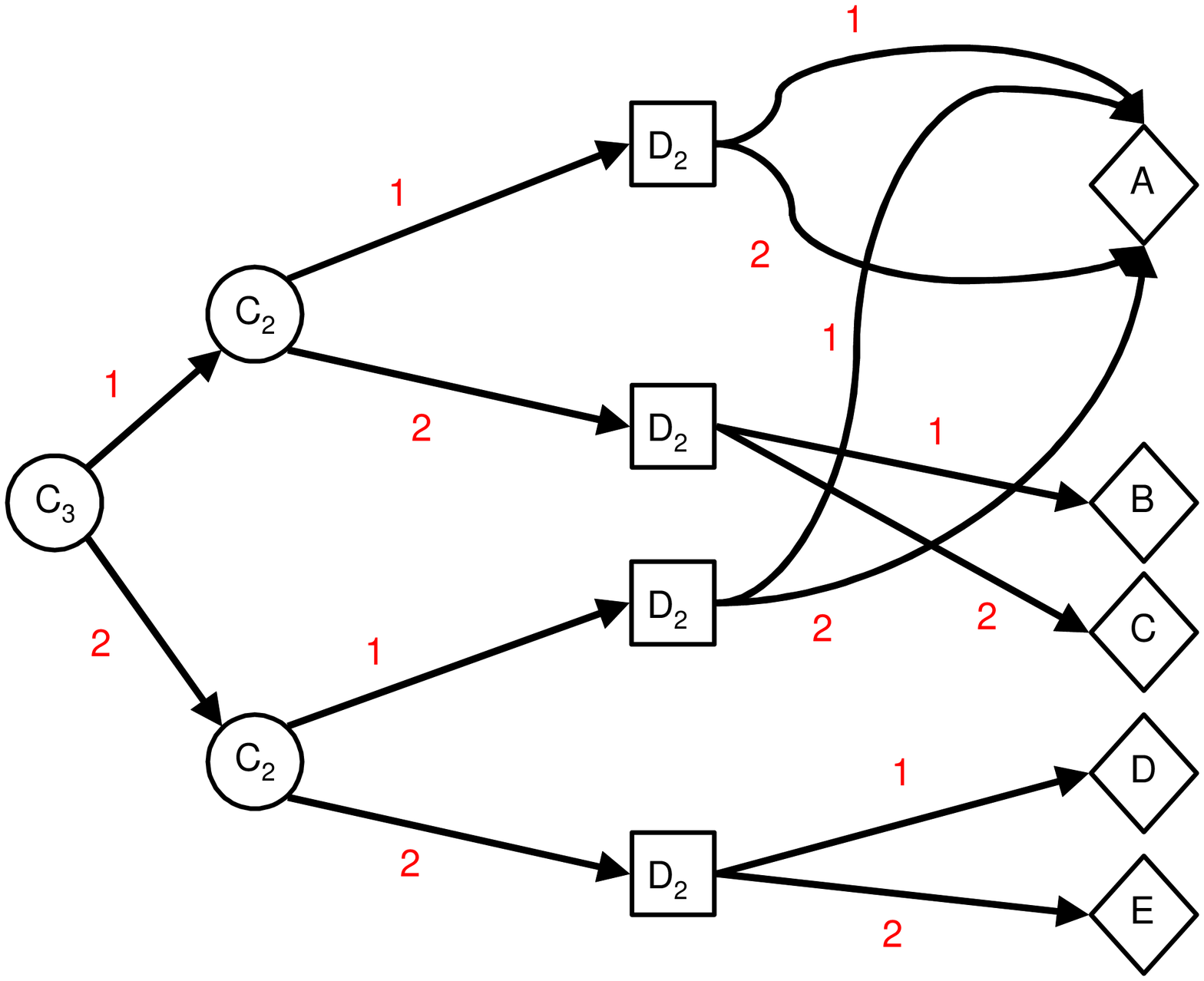}
\begin{center}
\caption{Distribution trees for nodes $C_1$ and $U$}
\end{center}
\vspace{-8mm}
\end{figure}

\smallskip
The ID in Figure 1 is not the most parsimonious representation of the problem.
If we can partition the parents of a decision
node $D$ (those nodes with arrows into $D$) into two sets $Q^A(D), Q^B(D)$ such that
$U \amalg Q^B(D)\ |\ (D, Q^A(D))$,
then the set $Q^B(D)$ can be considered {\it irrelevant} for the purposes of maximising utility, and the edges from nodes in $Q^B(D)$ into $D$ can be 
removed from the ID. Here we find that $C_1 \in Q^B(D_2)$, and so the edge from $C_1$ to $D_2$ can be removed from the ID. The node $C_1$ is now barren, 
so it can also be removed (together with the edge $D_1 \rightarrow C_1$). 

Once we have our parsimonious ID we can use one of the standard solution methods to produce an optimal decision strategy and expected utility
for this strategy. Using Shachter's method (reversing the arc between $C_3$ and $C_2$, and adding a new arc from $D_1$ to $C_3$) we eventually get
$$U^{final} = \max_{D_1}\Big[ \sum_{C_2}P(C_2\ |\ D_1)\Big[ \max_{D_2}\Big[ \sum_{C_3}P(C_3\ |\ D_1,C_2)\ U(C_2,C_3,D_2)\Big] \Big] \Big]$$
which does not however reflect the asymmetries in the problem. These can be built into the solution technique, but as the principal asymmetry concerns
$U(C_2,C_3,D_2)$, any advantage conveyed by the compactness of the ID is lost in the messy arithmetic.

\begin{figure}[ht]	
\begin{center}
\includegraphics[height=2.4in]{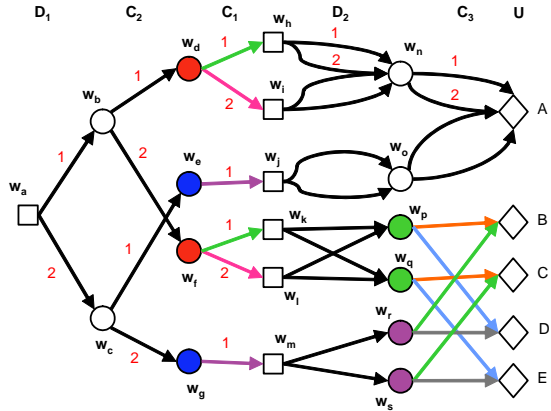}
\caption{Initial EF CEG for our example}
\end{center}
\vspace{-6mm}
\end{figure}

\smallskip
We now turn our attention to a CEG-representation of the problem. There are two EF orderings of the variables: 
$D_1,C_2,C_1,D_2,C_3,U$ and one where $C_1$ \& $C_2$ are interchanged. 
Note that $D_2$ precedes $C_3$ since the value of $C_3$ is not known to the DM when she comes to make a decision
at $D_2$. The first ordering leads to a slightly more transparent graph.

As we are comparing CEGs and IDs here, we do not put any utilities onto edges, but restrict them to terminal utility nodes. We also separate out 
our single utility node into distinct utility nodes for each 
value taken by $U$. In more complex decision problems this can lead to greater transparency. We have elsewhere called this form of CEG without 
utilities on edges, and with separated utility nodes, a Type~2 decision CEG.
The Type 2 CEG for the ordering $D_1,C_2,C_1,D_2,C_3,U$ is given in Figure~3.

\begin{figure}[ht]	
\begin{center}
\includegraphics[height=1.9in]{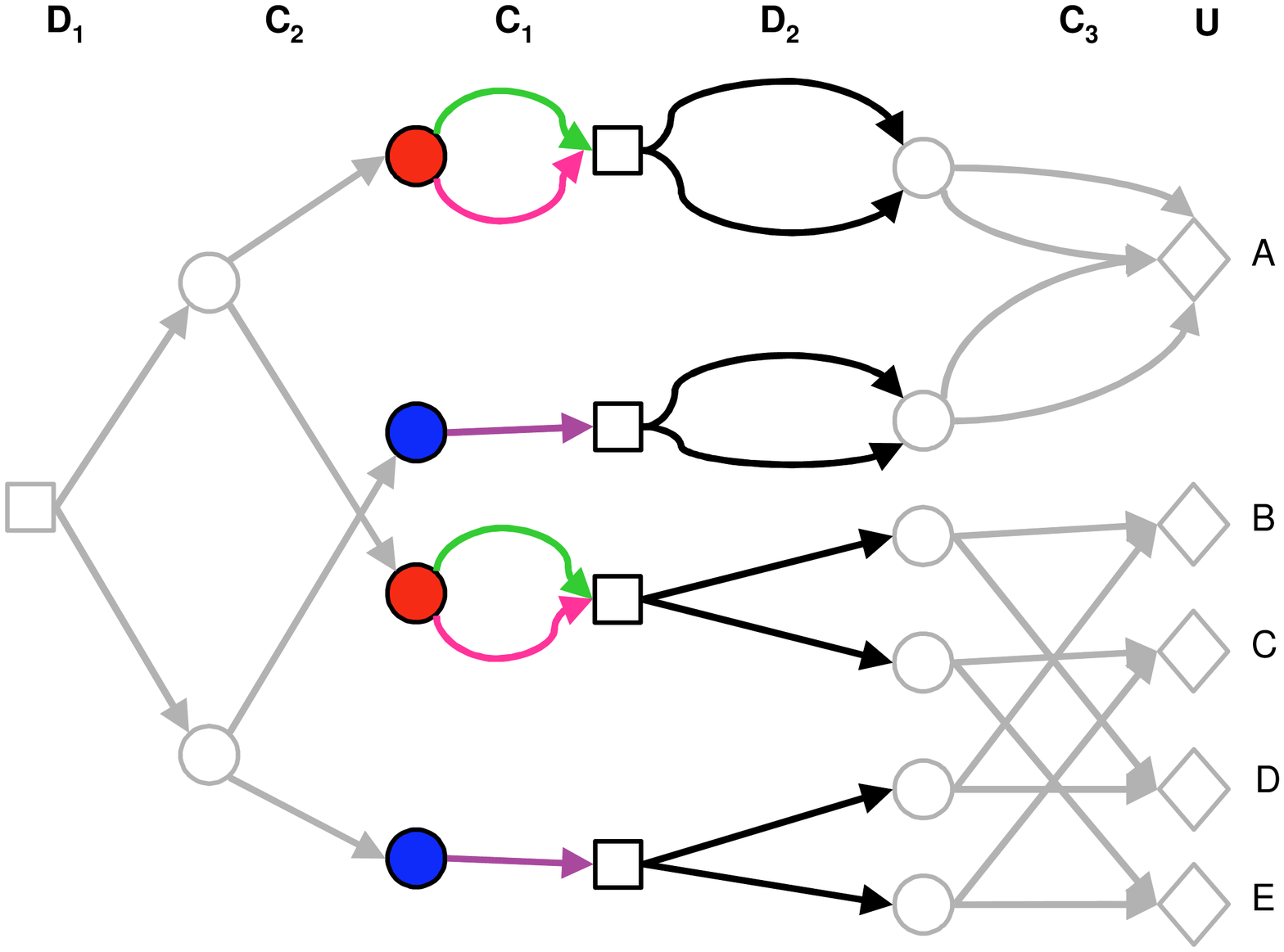}\kern-0.2in\includegraphics[height=1.9in]{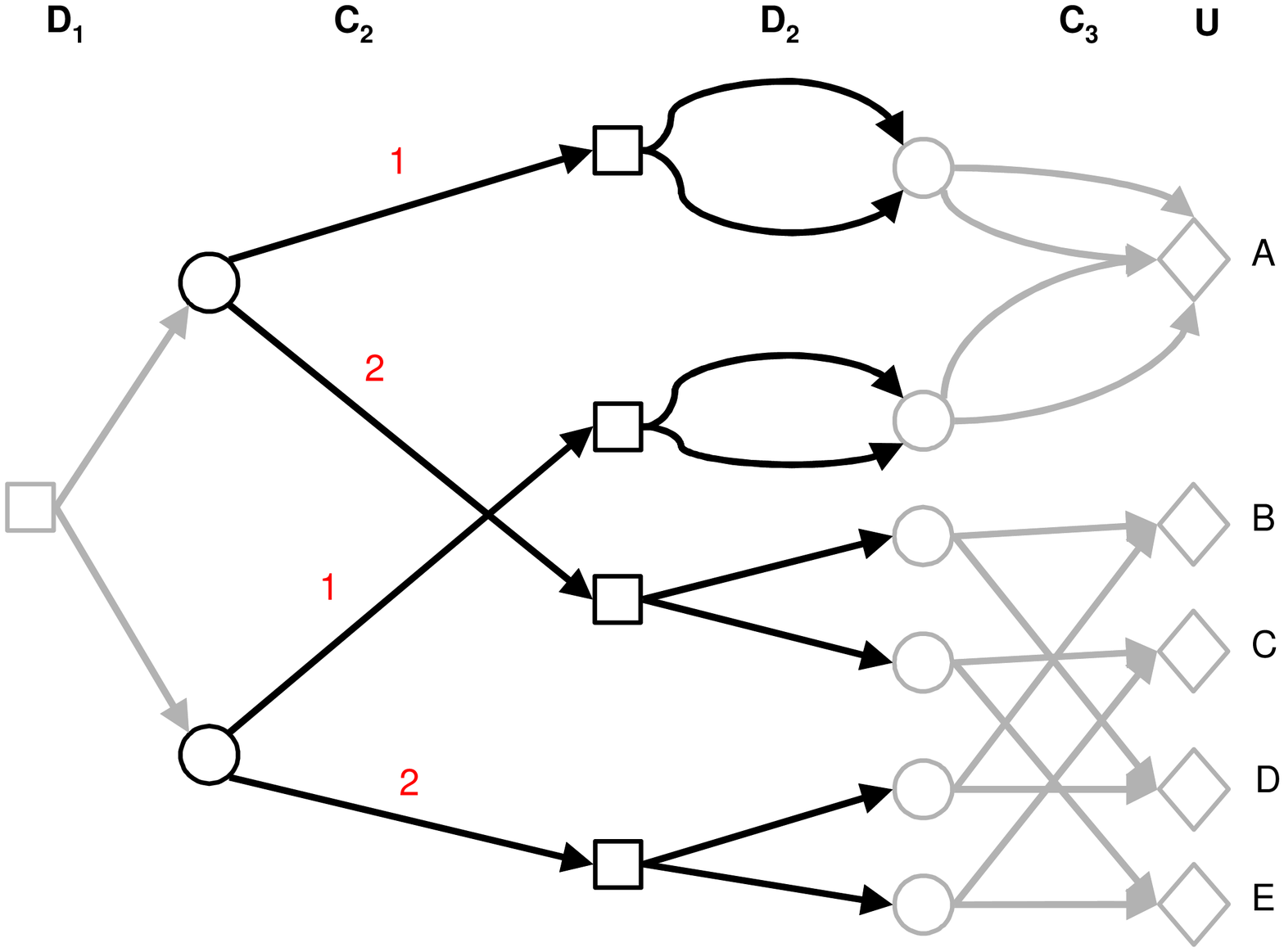}
\caption{First and second simplifications}
\end{center}
\vspace{-6mm}
\end{figure}

\smallskip
Conditional independence structure in a CEG can be read from individual positions, from stages, and from {\it cuts} through these~\cite{PaulandJim}.
Recall that nodes in the underlying tree are coalesced into positions when the sets of complete future developments from each node are the same and
have the same probability distribution. So for example, the position $w_n$ yields the information that
\begin{align*}
(C_3,U) &\amalg (C_1, D_2)\ |\ (D_1 = 1, C_2 = 1) &(3.1)
\end{align*}
The position $w_p$ similarly yields $(C_3,U) \amalg C_1\ |\ (D_1 = 1, C_2 = 2, D_2 = 1)$.

Recall that positions in a CEG are in the same stage if their sets of outgoing edges carry the same labels and have the same probability distribution. The positions
$w_p$ \& $w_q$ are in the same stage (indicated by the colouring), and so the probabilities on the edges leaving these positions have the same
distibution, and hence
\begin{align*}
C_3 &\amalg (C_1, D_2)\ |\ (D_1 = 1, C_2 = 2) &(3.2)
\end{align*}

The expressions (3.1) \& (3.2) result from the fact that in our EF CEG ordered $D_1,C_2,C_1,D_2,C_3,U$, the variable $C_3$ is dependent on $D_1$ and $C_2$. 
This is not clear from the ID in Figure~1, but is reflected in the expression for $U^{final}$. But the form of this expected utility expression
is a consequence of the arc-reversal required for successful ID-based solution of our problem. So this arc-reversal is already explicitly
represented in the original EF CEG, and is not (as with IDs) an additional requirement of the solution technique.

A {\it cut} through a CEG is a set of positions or stages which partitions the set of root-to-sink/leaf paths. So the set of positions $\{ w_n,w_o,
w_p,w_q,w_r,w_s \}$ is a cut of our CEG. A conditional independence statement associated with a cut is the union of those statements associated
with the component positions (or stages) of the cut. So the cut through $\{ w_n,w_o,w_p,w_q,w_r,w_s \}$ gives us that
\begin{align*}
U &\amalg C_1\ |\ (D_1, C_2, D_2) &
\end{align*}
which is clearly of the form $U \amalg Q(D_2^B)\ |\ (D_2, Q(D_2^A))$,
and tells us that $C_1$ is irrelevant to $D_2$ for the purposes of maximising utility.

For a Type 2 CEG drawn in EF order, two (or more) decision nodes are in the same position if the sub-CEGs rooted in each decision node have the
same topology, equivalent edges in these sub-CEGs have the same labels \& (where appropriate) probabilities, and equivalent branches terminate
in the same utility node. So in Figure~3, the nodes $w_h$ \&~$w_i$ are in the same position, as are the nodes $w_k$ \&~$w_l$.
Decision nodes in the same position can simply be coalesced, giving us the first graph in Figure~4.

For a Type 2 EF decision CEG with all positions coalesced (as in this graph), a barren node is simply a position $w$ for which $ch(w)$ 
(defined as in section~2) contains a single element. Barren nodes can be deleted in a similar manner to those in BNs -- see Table~3 (where
$pa(w)$ denotes the set of parent nodes of $w$).

\begin{table}[ht]		
\vspace{-4mm}
\caption{Barren node deletion algorithm (Type~2 decision CEGs) } 
\begin{center}
\begin{flushleft}
\begin{itemize}
\setlength{\itemsep}{-5pt}
\item Choose a topological ordering of the positions excluding the terminal utility nodes: $w_1, w_2,\ldots, w_m$, such that $w_1$ is the
  root-node.
\item Iterate: for $i=2$ step plus 1 until $i=m$ do:
\begin{itemize}
\item If $ch(w_i)$ contains only one node then\hfill\break
	Label this node $w_{\succ i}$\hfill\break
	For each node $w_{\prec i} \in pa(w_i)$\hfill\break
	Replace all edges $e(w_{\prec i},w_i)$ by a single edge $e(w_{\prec i},w_{\succ i})$\hfill\break
	Delete all edges $e(w_i,w_{\succ i})$ \& the node $w_i$.
\end{itemize}
\end{itemize}
\end{flushleft}
\end{center}
\vspace{-6mm}
\end{table}

Four iterations of the algorithm applied to the first graph in Figure~4 yield the second
graph in Figure~4. Further iterations will remove the first two $D_2$ nodes and the first two $C_3$ nodes to give the parsimonious CEG in Figure~5.

We can clearly see that $C_1$ is irrelevant for maximising $U$, and moreover if $C_2 = 1$ then both $D_2$ and $C_3$ 
are also irrelevant for this purpose (so the DM actually only needs to make one decision in this context). This latter property of the problem is not
one that can be deduced from an ID-representation, although it could with some effort be worked out from the second distribution tree in Figure~2.
It is however obvious in the parsimonious CEG. 

Solution follows the method described in section~2 (the process obviously being simpler as there are no rewards or costs on the edges), and results
in the expression
\begin{align*}
U^{final} &= \\
	\max\big[&P(C_2 = 1\ |\ D_1 =1) U_A + P(C_2 = 2\ |\ D_1 =1) \times \\
	\max\big[&P(C_3 = 1\ |\ D_1 = 1, C_2 =2) U_B + P(C_3 = 2\ |\ D_1 = 1, C_2 =2) U_D, \\
			&P(C_3 = 1\ |\ D_1 = 1, C_2 =2) U_C + P(C_3 = 2\ |\ D_1 = 1, C_2 =2) U_E \big], \\
		&P(C_2 = 1\ |\ D_1 =2) U_A + P(C_2 = 2\ |\ D_1 =2) \times \\
	\max\big[&P(C_3 = 1\ |\ D_1 = 2, C_2 =2) U_B + P(C_3 = 2\ |\ D_1 = 2, C_2 =2) U_D, \\
			&P(C_3 = 1\ |\ D_1 = 2, C_2 =2) U_C + P(C_3 = 2\ |\ D_1 = 2, C_2 =2) U_E \big] \big] 
\end{align*}

\begin{figure}[ht]	
\begin{center}
\includegraphics[height=2.4in]{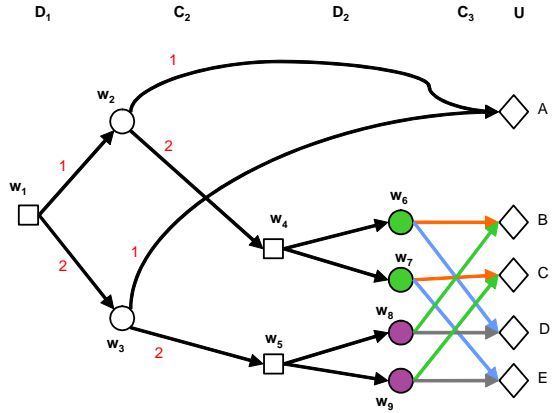}
\caption{Parsimonious CEG}
\end{center}
\vspace{-4mm}
\end{figure}

\noindent{This expression is obviously more complex than that for the ID, but it is much more robust since it has been produced using the asymmetry of 
the problem to power the analysis, rather than treating it as an added complication.} 

\section{Discussion}	

In this paper we have concentrated on how CEGs compare with IDs for the analysis of asymmetric decision problems. It is however worth pointing out
two advantages of CEGs over coalesced trees: Firstly, the ability to read conditional independence structure from CEGs allowed us to create an
analogue of the parsimonious ID, and secondly, the explicit representation of stage structure in CEGs gave rise to our barren node deletion
algorithm.

A paper providing a more detailed discussion of parsimony, barren node deletion and arc reversal as they relate to CEGs is imminent. This paper
will also provide a comparison of CEGs with VNs, SDDs and augmented IDs through a worked example. A further paper on the use of decision CEGs for 
multi-agent problems and games is also in the pipeline.

\bigskip
\noindent{{\bf Acknowledgement:} {This research is being supported by the EPSRC (project EP/M018687/1).}}


\begin{thebibliography}{99}

\bibitem{Lorna2}
Barclay L.~M., Hutton J.~L., and Smith J.~Q. (2013)
Refining a {B}ayesian {N}etwork using a {C}hain {E}vent {G}raph,
{\em International Journal of Approximate Reasoning}, 54:1300--1309.

\bibitem{BhatandS2}
Bhattacharjya D. and Shachter R.~D. (2012)
Formulating asymmetric decision problems as decision circuits,
{\em Decision Analysis}, 9:138--145.

\bibitem{BandShen}
Bielza C. and Shenoy P.~P. (1999) 
A comparison of graphical techniques for asymmetric decision problems,
{\em Management Science}, 45:1552--1569.

\bibitem{CandM}
Call H.~J. and Miller W.~A. (1990)
A comparison of approaches and implementations for automating {D}ecision analysis,
{\em Reliability Engineering and System Safety}, 30:115--162.

\bibitem{CovandOli}
Covaliu Z. and Oliver R.~M. (1995)
Representation and solution of decision problems using sequential decision diagrams,
{\em Management Science}, 41(12).

\bibitem{Dawid1979}
Dawid A.~P. (1979)
Conditional independence in statistical theory,
{\em Journal of the Royal Statistical Society, Series B}, 41:1--31.

\bibitem{DawidinnewCausalbook}
Dawid A.~P. (2012)
The decision-theoretic approach to causal inference.
In C.~Berzuini, A.~P. Dawid, and L.~Bernardinelli, editors, {\em Causality: Statistical Perspectives and Applications}, pages 25--42. Wiley.

\bibitem{JNandShen}
Jensen F.~V., Nielsen T.~D., and Shenoy P.~P. (2006)
Sequential influence diagrams: A unified asymmetry framework,
{\em International Journal of Approximate Reasoning}, 42:101--118.

\bibitem{QZP}
Qi R., Zhang N., and Poole D. (1994)
Solving asymmetric decision problems with influence diagrams.
In {\em Proceedings of the 10th Conference on Uncertainty in Artificial Intelligence}, pages 491--499.

\bibitem{Shachter86}
Shachter R.~D. (1986)
Evaluating {I}nfluence diagrams,
{\em Operations Research}, 34(6):871--882.

\bibitem{Shenoy96}
Shenoy P.~P. (1996)
Representing and solving asymmetric decision problems using valuation networks.
In D.~Fisher and H-J. Lenz, editors, {\em Learning from Data: Artificial Intelligence and Statistics V}. Springer-Verlag.

\bibitem{SilanderCEG}
Silander T. and Leong T-Y. (2013)
A {D}ynamic {P}rogramming {A}lgorithm for {L}earning {C}hain {E}vent {G}raphs.
In {\em Discovery Science}, volume 8140 of {\em Lecture Notes in Computer Science}, pages 201--216. Springer.

\bibitem{SHM}
Smith J.~E., Holtzman S., and Matheson J.~E. (1993)
Structuring conditional relationships in influence diagrams,
{\em Operations Research}, 41:280--297.

\bibitem{PaulandJim}
Smith J.~Q. and Anderson P.~E. (2008)
{C}onditional independence and {C}hain {E}vent {G}raphs,
{\em Artificial Intelligence}, 172:42--68.

\bibitem{newcap}
Thwaites P.~A. (2013)
Causal identifiability via {C}hain {E}vent {G}raphs,
{\em Artificial Intelligence}, 195:291--315.

\bibitem{CausalAI}
Thwaites P.~A., Smith J.~Q., and Riccomagno E.~M. (2010)
Causal analysis with {C}hain {E}vent {G}raphs,
{\em Artificial Intelligence}, 174:889--909.

\end{thebibliography}

\end{document}